\documentclass[trackchanges,twocolumn]{aastex702}

\def\gtorder{\mathrel{\raise.3ex\hbox{$>$}\mkern-14mu
    \lower0.6ex\hbox{$\sim$}}}
\def\ltorder{\mathrel{\raise.3ex\hbox{$<$}\mkern-14mu
    \lower0.6ex\hbox{$\sim$}}}

\newcommand{\ZZ}[1]{}  

\usepackage{graphicx}	
\usepackage{cancel}     
\usepackage{amsmath}	
\usepackage{soul}
\usepackage{multirow}
\usepackage{hhline}

\begin{document}

\title[Self-gravitating Disks]{The Limited Lifetime of Self-Gravitating Accretion Disks in Galactic Centers}

\author[0000-0002-1233-445X]{Isaac Shlosman}
\affiliation{Department of Physics \& Astronomy, University of Kentucky, Lexington, KY 40506, USA}\email{isaac.shlosman@uky.edu}
 
\correspondingauthor{Isaac Shlosman}
\email{isaac.shlosman@uky.edu}

 \begin{abstract}
Globally self-gravitating accretion disks in active galactic nuclei (AGN) have been frequently discussed in the literature, but never observed. Few dozen megamasering accretion disks in AGN have been detected, which display a clear Keplerian rotation on scales of 0.1-1\,pc, allowing to estimate their masses within the sphere of influence (SoI) of the central supermassive black holes (SMBHs) to be smaller by more than a factor of 10, compared with the parent SMBH masses. Furthermore, the stellar components, deep inside the SoI, are dwarfed by the SMBH masses of $M_\bullet\sim {\rm few}\times 10^6 - {\rm few}\times 10^7\,M_\odot$. Theoretically, no limit exists on sizes and masses of gaseous disks in AGN, which in principle, can exceed masses of their central compact objects. We analyze instabilities which can operate in gaseous disks, i.e., fragmentation for a locally dominant self-gravity and spontaneous breaking of axial symmetry for the globally self-gravitating disks. We invoke the gas response to the latter instability which leads to gravitational collapse, leaving a negligible mass behind and dynamically stable remnants. Consequently, the characteristic timescale for globally self-gravitating disks to exist in AGN should not exceed few rotations, $t_\phi\sim 10^{3-4}$\,yrs for $M_\bullet\sim 10^{6-8}\,M_\odot$. Disk rebuilding is expected to be $\gtorder 10^7$\,yrs, meaning that probability of finding is $\ltorder 10^{-3}$, which explains their lack of detection. We conclude that observed sub-parsec disks in AGN, at least the Keplerian masering ones, can be remnants of globally self-gravitating accretion disks, and the instability timescale can be related to the duty cycle of AGN, $t_{\rm duty}\sim 10t_\phi\sim 10^{4-5}$\,yr.  
 \end{abstract}


\section{Introduction}  
\label{sec:intro}

Highly inclined masering accretion disks, observed through 22\,GHz, have been detected in the central parsec of active disk galaxies \citep[e.g.,][]{miyoshi95} --- in Seyfert\,2 and Low-Ionization Nuclear Emission (LINER) galaxies. Within the specific physical conditions in these objects, such as the temperature and density, water molecules are collisionally pumped to emit highly focused, amplified microwave radiation, known as water megamasers. The measured rotation appears to be Keplerian, to better than 1\% \citep[e.g.,][]{modjaz05}, allowing to obtain the masses of the central supermassive black holes (SMBHs), $M_\bullet\sim {\rm few}\times 10^6 - {\rm few}\times 10^7\,M_\odot$ \citep{kormendy13}, constrain the Hubble constant \citep[][and refs. therein]{pesce20}, and provide important information about low-mass end of the $M_\bullet-\sigma_*$ relation, where $\sigma_*$ is the effective stellar dispersion velocity in the bulge. In addition to determining the BH demographics at various redshifts, it enables to estimate the masses of accretion disks, which appear smaller by a factor larger than 10 in comparison to $M_\bullet$ \citep[e.g.,][]{kuo18}.   

Masses of the SMBHs have been estimated in few dozen AGN accretion disks with megamasers, relying on the Very Long Baseline Interferometry (VLBI) and ALMA \citep[e.g.,][and refs. therein]{lo05,genzel10,kuo25}. Stellar dynamics methods provide a lower accuracy, due to the unknown orbital anisotropy, dark matter (DM) contribution, and converting light to mass distribution, including the initial mass function (IMF).  Using the water megamasers provides an alternative and promising approach to reduce the uncertainties in $M_\bullet$ measurements, and favor Type\,II AGN which harbor nearly edge-on disks on scales of 0.1-1\,pc, allowing the VLBI to identify masering spots with $\sim 10\,\mu$as precision \citep[][]{gao17}. The inner edge of the masering disks have been projected to lie around the dust sublimation radii \citep[e.g.,][]{kuo24}, and therefore can coincide with the outflow region referred to as the NIR torus \citep{elitzur06}. This warm gas with $T\sim 100-1,000$\,K probes kinematic and spatial distributions of H$_2$ within the radius of the sphere of influence of the SMBH, defined\footnote{ Note that two definitions of the sphere of influence for the SMBH exist: (1) based on the stellar dispersion velocities, $GM_\bullet/\sigma_*$, and (2) based on the enclosed mass, $2M_\bullet $. These definitions can differ by about a factor of 2 -- 3.} here by $R_{\rm h} \sim 4.1\,{\rm pc}\, (M_\bullet/10^7\,M_\odot)/(\sigma_*/200\,{\rm km\,s^{-1}})^2$, where $\sigma_*$ is stellar dispersion velocity. High gas densities in the range of $\sim 10^7-10^{11}\,{\rm cm^{-3}}$ are required \citep{elitzur89,kuo25}.

The estimated masses of accretion disks in masering AGN, being limited by a factor of 10 below the central SMBHs, is puzzling. Classical accretion $\alpha$-disks \citep[][]{shakura73} and other disk models have no obvious limits on their masses or sizes, depend on the mass accretion rates, and, in principle, can exceed masses of the parent SMBHs. 

In this paper, we analyze properties of these disks, providing a plausible  explanation to the absence of massive accretion disks, approaching or even exceeding their central compact objects. Section\,\ref{sec:distrib} presents the inferred stellar and gas distributions in the AGN. Section\,\ref{sec:disks} is devoted to properties of massive gaseous disks and related instabilities, followed by Discussion section.

\section{Stellar and gas distributions around SMBHs}
\label{sec:distrib}

SMBHs are frequently embedded in stellar cusps with a density $\sim R^{-\gamma}$, which are more compact around disk galaxies and low mass ellipticals, and less compact around massive elliptical galaxies. The best studied Milky Way (MW) stellar cusp ($1" = 4\times 10^{-2}$\,pc, corresponding to the distance of 8.25\,kpc), has most of the stars inside the central pc older than 1\,Gyr. They display an isotropic distribution which slowly rotates in the sense of the MW. Young stars appear to populate a thin inclined disk to the MW plane (or two disks, \citet{tanner06}), with an average orbital eccentricity of 0.3, and are on the near-Keplerian orbits \citep{rauch96}. The stellar density distribution, has a break in the power law of $\sim R^{-1.3\pm 0.1}$ at $R\sim 0.25$\,pc, which steepens outside to $\gamma\sim 1.8$ \citep{genzel10}. The enclosed cluster mass within $5"$, i.e., 0.2\,pc, is dominated by $M_\bullet$ \citep{gravity22}, in agreement with \citet{bartko10} within 0.5\,pc.  

More generally, theoretical estimates of stellar distribution around the SMBHs resulted in pure stellar cusps with a range of slopes, $\gamma\approx 1.75$ \citep{bahcall76,frank76}, and $\gamma\approx 1.5$ \citep{young80}.  Generalized solutions have been introduced \citep{lightman77}, including nuclear star clusters with a substantial gas component which can be retained in more massive clusters \citep{krause16}. In the presence of stellar collisions, the slope becomes substantially shallower, down to $\gamma\sim 0.5$ \citep{murphy91}. Deviations from Bahcall-Wolf slope are expected if the cluster did not reach equilibrium after a strong perturbation by a penetrating merger and in-spiraling SMBH \citep{baumgardt06}. 

The sphere of influence of central SMBHs has been estimated to be in the range of a few pc --- few tens of pc, for the $M_\bullet$ range discussed in this work. Therefore, the masering accretion disks are found deep inside the sphere of influence of the SMBH, and hence the stellar component at these radii is completely dwarfed by the SMBH mass \citep[e.g.,][]{schartmann10,pechetti20}. While this appears to be well established for $M_\bullet > 10^6\,M_\odot$, smaller SMBHs, if they exist, may have nuclear star clusters dominating the SMBH mass. But such AGN lie outside the mass range of detected megamasering accretion disks.

We turn now to gas distribution within central parsec of the AGN in general and masering AGN in particular. In a carefully studied NGC\,4258, the maser emission displays a perfect Keplerian rotation, $r^{-0.5}$, in a geometrically-thin, cool and dense disk. Its masering disk lies in the range of 0.13-0.26\,pc \citep[e.g.,][]{moran95} and the disk mass was estimated to be $M_{\rm disk}\ltorder (1-2)\times 10^5\,M_\odot$ \citep{miyoshi95,lo05} for $M_\bullet \approx 4.08\times 10^7\,M_\odot$ \citep{nguyen26}. An exhaustive analysis of the gas kinematics has rejected a more massive disk possibility in favor of a warped accretion disk \citep{herrnstein05}.  

Basically all the H$_2$O masering disks display a remarkable Keplerian rotation curves as they probe the gas motion inside the sphere of influence of the central SMBH of $M_\bullet \sim 2\times 10^6 - 6\times 10^7\,M_\odot$. While a number of thermodynamical parameters describe the masering gas, the most important appears the H$_2$ gas temperature in the range of $400\ltorder T \ltorder 1500$\,K and density $10^7-10^{11}\,{\rm cm^{-3}}$ \citep[e.g.,][and refs. therein]{kuo25}. In addition, the inferred median Eddington ratio of $\sim 0.04$ for these AGN can reflect the stable molecular gas properties in these objects. 

The non-Keplerian rotation has been claimed for accretion disk in NGC\,1068, with $r^{-0.31}$ out to 1\,pc \citep{greenhill96}, seconded by a measurement method described in \citet{hure11}. Using this analysis, \citet{lodato03} suggested that the mass of the SMBH in this galaxy is of the order of the accretion disk mass. However, this result of Lodato and Bertin for NGC\,1068 was re-analyzed and disputed by \citet{kuo18}, obtaining the disk mass at least 10 times smaller than its parent SMBH. New observations of this galaxy agree with a Keplerian rotation in $M_{\rm disk}\sim 1.1\times 10^5\,M_\odot$ accretion disk around $M_\bullet\sim 1.7\times 10^7\,M_\odot$ \citep{gallimore23}. Moreover, recent detection of a megamaser disk in NGC\,7738 with a claimed non-Keplerian rotation \citep{ito25} has been refitted and found to be consistent with Keplerian rotation \citep{hopkins26}.

In summary, the water maser disks existing between 0.1-1\,pc around the SMBHs of $\sim 10^7\,M_\odot$ display a Keplerian rotation. The disk masses within the sphere of influence of the SMBH are more than a factor of 10 smaller than the parent $M_\bullet$. Therefore, it is safe to conclude that these disks are not found in the regime when they are dominated by self-gravity, especially the global self-gravity. As there are no obvious limits on the accumulated amount of gas near the central SMBH, it is surprising that more massive self-gravitating disks are not observed in these objects. In the next section we analyze the evolution of such disks and come to the conclusion that the apparent absence of such disks in AGN is not accidental, but is a direct result of their evolution. 

\section{Local and global instabilities in self-gravitating accretion disks}
\label{sec:disks}

Dynamics of self-gravitating fluids has been studied since Isaac Newton \citep[][and refs. therein]{hunter72}. It has been analyzed in the context of accretion onto the SMBHs in AGN \citep[e.g.,][]{shlosman87,shlosman89a,shlosman89b,bertin97,lodato08}, and the mass accumulation in the centers of DM halos leading to the formation of Pop\,III stars and SMBH seeds \citep[e.g.,][]{begelman06,wise08,bromm09,begelman09}. 

Self-gravitating disks are subject to two important dynamical instabilities, i.e., local and global ones. The local (Jeans) instability  leads to disk fragmentation and under certain conditions to star formation. The characteristic scale of this instability is much smaller than the disk size, $\lambda << R_{\rm disk}$, which can be translated to $M_{\rm disk}/M_\bullet << 1$. 

The global instability is related to the spontaneous symmetry breaking in the disk flow, and is dominated by the $m=2$ Fourier global mode, i.e., by the gaseous (or stellar) bar mode. Global self-gravity confines the disk in the radial direction, and the related instability has $\lambda\sim R_{\rm disk}$. For this reason it is difficult to treat analytically, except in some special cases, because the WKB approximation cannot be applied, and because global modes cannot be captured using shearing box simulations. This instability has a dramatic effect on the radial gas and angular momentum flow, has been discovered theoretically and modeled numerically. For the purpose of this work the local gravitational instability has only a peripheral interest, and is discussed below in order not to mix it with the global gravitational instability.

The difference between both types of disks is fundamental: the Jeans instability depends on the local surface density and its cooling time, and the disk remains in a Keplerian rotation. On the other hand, a global self-gravity unavoidably modifies the mass distribution, and the circular velocity $v_{\rm c}^2=GM(<R)/R$ is not Keplerian and does not follow the $R^{-1/2}$ law. Jeans-unstable disks have been frequently invoked in this context \citep[e.g.,][]{shlosman87,shlosman89a,gammie01,hopkins26}. Globally self-gravitating gaseous disks have never been observed in AGN, but were analyzed in this context \citep[e.g.,][]{shlosman89b,begelman09}. Hence, it is naturally to ask if they avoided detection so far, or some physics prevents their formation, or their lifetime is short, which again complicates their detection.

Additional important point: in this work we focus on the evolution of gaseous self-gravitating disks in the absence of dynamically important magnetic fields. Although magnetic fields are known {\it theoretically} to be amplified in accretion flows around compact objects \citep{begelman23a,begelman23b,hopkins24}, the observational evidence is only circumstantial. If their existence in AGN is confirmed, they can resolve some issues related to the fragmentation in accretion disks by means of magnetic torques. However, gravitational torques can resolve these evolutionary issues as well --- here we explore this alternative.

The outer part of $\alpha_{\rm ss}$-disks \citep{shakura73} is associated with physical processes governed by self-gravitational effects.  Accepting the original $\alpha_{\rm ss}$-disk model prescription with the surface density of $\Sigma\sim r^{-3/4}$ in the outer disk\footnote{Here $r$ corresponds to the cylindrical radius.}, and with a thermal pressure and free-free opacity, would lead to a rotation curve based on the mass distribution $M_{\rm disk}(<r)\sim r^{5/4}$. Such a disk dominates in mass, and competes with the surrounding stellar mass on scales of $\ltorder 1$\,pc. Moreover, evolution of these disks depends on degree to which self-gravity dominates, as we discuss below. 

A cooling disk with a scale height $h$ is subject to Jeans instability, when its surface density surpasses the critical value --- the so-called vertically self-gravitating disk \citep{toomre64}. The resulting disk fragmentation, i.e., the Jeans instability, depends crucially on the mechanism supporting the vertical thickness of the disk, which is characterized by max$(c_{\rm s},v_{\rm turb})$, where $c_{\rm s}$ and $v_{\rm turb}$ are the sound and turbulent velocities. Disk fragmentation can be triggered with accretion rate in excess of $\dot M\sim 3\alpha_{\rm ss}\,{\rm max}(c_{\rm s},v_{\rm turb})^3/G$ \citep{shlosman89a}, where $\alpha_{\rm ss}$ is the \citet{shakura73} viscosity parameter. Star formation with yet undetermined efficiency is a direct result of this instability \citep[e.g.,][]{shlosman89a,gammie01}.
 
A more important instability for our discussion is the global dynamical instability in a self-gravitating gaseous disk. Such a disk is confined radially by its own gravity. This instability results from a spontaneous symmetry breaking, so-called bar instability. The critical parameter which determines the onset of global instability is the ratio of rotational kinetic energy to the absolute value of gravitational potential energy, T/$|$W$|$, in the disk, which is the main tool used to analyze the global stability of such systems. For incompressible Maclaurin spheroids, the dynamical instability develops for T/$|$W$|\ge 0.2738$ \citep[e.g.,][and refs. therein]{chandra69}. This result has been confirmed approximately for a long list of models, independently of a polytropic index, distribution of angular momentum, and nonuniform densities \citep[e.g.,][]{ostriker73,bardeen75,tohline85}, confirming that instability is dynamical. The instability develops on the local free-fall or rotational timescale.

An alternative and physically justifiable derivation of the global instability in fluid systems has been proposed, based on the angular momentum content rather than on the energy content \citep{christo95b,christo95a}. It was applied successfully to axisymmetric models with different geometries, differential rotation and nonuniform density distributions. These systems included gaseous Maclaurin disks \citep{hunter79}, gaseous spheroids \citep{ostriker73,tohline85}, and gaseous tori \citep{woodward94}. The non-axisymmetric systems have been analyzed in \citet{christo95c}.

The condition for appearance of non-axisymmetric $m=2$ Fourier (bar) modes due to a dynamical instability in uniformly or differentially rotating gaseous axisymmetric systems is given by \citet{christo95b},
\begin{equation}
\alpha\equiv \bigg(\frac{1}{2}\, f\, \frac{T}{|W|}\bigg)^{1/2} > \alpha_{\rm crit} \approx 0.35,
\end{equation}
where $0\le \alpha \le 1/2$, $f$ is the shape factor, and $f=1$ for disks and $f=2/3$ for spheres. More generally, $f=(2/3)(1+T/|W|)$ for a thick, differentially rotating spheroidal mass distribution. The shape factor depends on both geometry and topology of the system \citep[e.g.,][]{tohline90}. For gaseous systems,  the dynamical bar instability is triggered spontaneously at $\alpha_{\rm crit}\approx 0.34-0.35$, so with an uncertainty of $\sim 2.9\%$. 

The physical nature of the $\alpha$ parameter becomes more transparent if it is written in terms of the disk angular momentum $J$, namely,
 
\begin{equation}
\alpha = \frac{5}{4}\, f\, \frac{J/M_{\rm disk}}{\Omega_{\rm J} a_1^2},
\end{equation}
where $M_{\rm disk}$ is the disk mass, $a_1$ is its equatorial radius, and $\Omega_{\rm J}$ is the Jeans radial frequency,

\begin{equation}
\Omega_{\rm J}^2 = 2\pi G <\rho> A(f), 
\end{equation}
with  the average mass density $<\rho>$,  and $A(f)$ being a function of the shape factor. Note that $5f/4\approx 1$ for spheroids and $f=1$ for disks. The product $\Omega_{\rm J}a_1^2$ is the maximal angular momentum of a circular orbit in the midplane.

What instability acts first, the local or global one?  This depends on the heating and cooling processes in accretion disks. If heating is dominated by viscous dissipation in differentially rotating disks, then cooling can drive the disk to Jeans instability which depends on the restoring pressure. In the absence of magnetic fields, it can be either thermal or turbulent pressure. Turbulence can be driven by the infall which releases gravitational energy, and by the magneto-rotational instability (MRI, \citet{balbus91}).  In a razor-thin disk, the dispersion relation, $\omega(k)^2=\kappa^2-2\pi G\Sigma|k| +c_{\rm s}^2k^2$, which defines the stability parameter $Q=\kappa c_{\rm s}/\pi G \Sigma$ \citep{toomre64} against fragmentation.  Here $k$ is the wavevector, and $\omega(k)$ and $\kappa$ are the wave and epicyclic frequencies, respectively, and $c_{\rm s}$ is the sound speed \citep[see also][]{binney08}. The Toomre parameter, $Q$, provides a boundary for stability. The locally self-gravitating disk will settle into a steady state with a fixed $Q$ \citep[e.g.,][]{paczynski78,gammie01}. 

For a more realistic 3-D gaseous disk, a finite thickness, $h$, should be introduced into dispersion relation by solving the Poisson equation for a vertically exponential disk \citep{begelman09},   
\begin{equation}
   \omega(k)^2 = \kappa^2 - \frac{2\pi G\Sigma |k|}{1+|k|h}  + v_{\rm turb}^2k^2,
\end{equation}
thus reducing the value of destabilizing gravity term by accounting for `puffing up' the disk by turbulence. Note, that while the local (Jeans) instability depends crucially on the thermal state of the gas, the global instability depends on the ratio of rotational kinetic energy to gravitational potential energy (Eqs.\,1 and 2). 

The ratio T/$|$W$|$ can be expressed in terms of rotational-to-Keplerian velocity ratio, $v_\phi/v_{\rm k}$, and $\alpha_{\rm crit}$ (Eq.\,1) corresponds to $v_\phi/v_{\rm k}\approx 0.57$ \citep{begelman09}. For a disk thickened by a fully developed turbulence, the global instability is triggered well before fragmentation sets in. Indeed, numerical simulations have shown the development of gaseous bars and absence of fragmentation during a runaway infall \citep[e.g.,][]{choi13}.

Angular momentum transport in a locally self-gravitating systems is due to tightly wound spiral density waves \citep[e.g.,][]{lynden72,lin87,laughlin98}. However, this mechanism originates from a local instability and, therefore, is entirely local in nature \citep[e.g.,][]{lodato04}, although for the $M_{\rm disk}/M_\bullet \rightarrow 1$, the {\it nonlocal} energy and angular momentum transfer dominate \citep{balbus99,cossins09,forgan11}. 

Importantly, local stability of self-gravitating disks does not ensure their global stability, as it completely ignores the modes which are not tightly wound. It even does not distinguish between axisymmetric and non-axisymmetric perturbations when they are tightly wound.

In a globally self-gravitating disk, the effective angular momentum transfer due to gravitational torques is non-local in nature, and couples disk regions on the scale of a disk size by developing the bar mode. It, therefore, triggers large scale gas inflow \citep{shlosman89b,shlosman90,lodato05}. Growing chaotic motions in a heavily non-axisymmetric disks, are accompanied by shocks. This evolution cannot proceed in the stellar systems, which increase their dispersion velocity when contracting, terminating the inflow.

To demonstrate how the disk stability depends on the $M_{\rm disk}/M$ ratio in the range of 0.125-0.5, where $M$ is the central mass, we refer to a qualitative comparison with modeled proto-planetary disks \citep[e.g.,][]{dong15}. Grand-design spiral arms become much more prominent and the $m=2$ mode starts to dominate. This affects not only the disk structure, but substantially increases the mass accretion rate and the rate of the angular momentum transfer.

For the sake of simplicity, to demonstrate the action of radial inflow due to the global instability, we use the model evolved in \citet{englmaier04} of a disk galaxy with the  \citet{miyamoto75}
potential, which describes the DM halo with embedded stellar disk and bulge. Using the final stage of the numerical simulation, we focus on a self-gravitating gas disk accumulating inside the central kpc, and describe the disk evolution after it reached about $\alpha_{\rm crit}$ (Eq.\,1), forming a gaseous bar, triggering  internal shocks, and a catastrophic loss of the angular momentum.  

During the collapse, the gaseous bar increases its ellipticity up to $e\sim 0.7-0.8$, i.e., this is a strong bar. Its semi-major axis shrinks from $a\sim 1$\,kpc by a factor of 10, terminated only by the finite numerical resolution, when the bar dissolves at $a\sim 100$\,pc. The gas inflow rate increases dramatically by more than a factor 10 over time period of $\sim 100$\,Myr, declining thereafter with the bar dissolution. During this time, the bar pattern speed, $\Omega_{\rm bar}$, increases by a factor of 9, and was shown to follow $\Omega_{\rm bar}\sim a^{-1}$. 

The characteristic rise time of instability inside $a\sim 1$\,kpc using orbital velocity (in the above example), $v_\phi\sim 100\,{\rm km\,s^{-1}}\sim (G M_{< 1\,{\rm kpc}}\,/a)^{1/2}$, where $M_{\rm <{\rm 1\,kpc}}$ is the enclosed mass within 1\,kpc, has been compared to the orbital time at the bar radius, $t_\phi\sim 2\pi a/v_\phi\sim 63\,a_{\rm kpc}/v_{\rm \phi,100}$\,Myr. The about axisymmetric gas disk has formed a strong bar with nearly all the gas concentrated within the bar, followed by the inflow, which is essentially an avalanche. The amount of gas remaining behind is negligible, and its surface density has decreased by orders of magnitude. 

Such a cascade can be maintained over a substantial dynamic range in radius and can be related to self-organized criticality \cite[][]{licht92}. An analogy can be made between the chaos driven when the bar exceeds a certain strength and a sandpile whose slope is increased until the sand slides off. This model exemplifies that globally self-gravitating gaseous disks self-destruct by channeling the gas inwards, until the conditions supporting the existence of these configurations have ceased. This process of a catastrophic inflow is suitable for dissipative systems only. The disk regains stability by substantially decreasing its surface density in the outer self-gravitating region.

In physical systems, self-organized criticality describes how these systems naturally evolve into an unstable state. Evolving by a slow input of mass or energy, such systems release stress through cascading avalanches. The self-organized criticality creates scale-free, power-law distributions without requiring fine tuning from outside forces.

Important conclusion we obtain from this representative example is that globally self-gravitating gaseous disks in galactic centers are unstable and exist during a limited time only, i.e., about few orbital timescales in the outer disk radii, where most of the gas has accumulated. Extrapolating down to $\sim 0.1$\,pc accretion disk, the characteristic timescale for instability becomes 

\begin{equation}
t_\phi\sim 4.2\times 10^2 \bigg(\frac{\delta}{5}\bigg)^{-1/2}\bigg(\frac{r}{0.1\,{\rm pc}}\bigg)^{3/2}\bigg(\frac{M_\bullet}{10^7\,M_\odot}\bigg)^{-1/2}\, {\rm yrs}.
\label{eq:5}
\end{equation}
Here, we have introduced a fudge factor $\delta\sim 3-10$, which describes the mass of a self-gravitating disk, $\sim \delta M_\bullet$, prior to the onset of instability. It is difficult to estimate $\delta$ more precisely, as it depends on accretion rate onto the disk.

\begin{figure}[ht!]
\center
\includegraphics[width=0.8\linewidth]{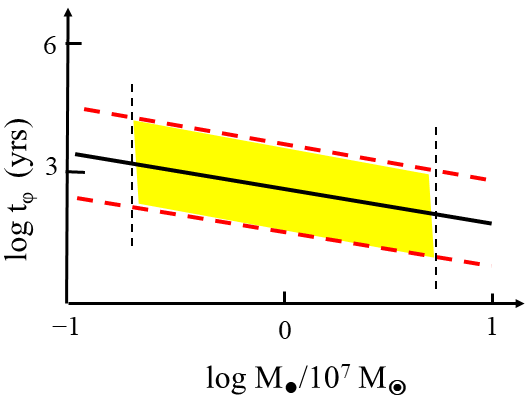}
\caption{Characteristic timescale for gaseous bar instability, $t_\phi$, given by equation\,\ref{eq:5},  $t_\phi\sim (r/0.1\,{\rm pc})^{3/2} (M_\bullet/10^7\,M_\odot)^{-1/2}$, with the fudge factor $\delta=5$. The black solid line corresponds to $r=0.1$\,pc and $M_\bullet=10^7\,M_\odot$. The red dashed lines correspond to $r=10^{-2}$\,pc (bottom) and $r=1$\,pc (top). Vertical dashed lines correspond to the observed range of $M_\bullet\approx 2\times 10^6\,M_\odot - 6\times 10^7\,M_\odot$ in megamasering accretion disks of AGN.
\label{fig:time}}
\end{figure}

Taken conservatively, this results in $t_\phi\sim 10^{3-4}$\,yr, meaning that the disk can exist a very short time in its massive state (see also Figure\,\ref{fig:time}), compared to the dynamical time which characterizes processes in the galactic disk on scales of $\sim 1-10$\,kpc. Here, we do not address the fate of this gas inflow, which can lead either to a powerful starburst or feeding the SMBH accretion. The characteristic timescale $t_\phi$ can be rather related to the triggering the active phase of the AGN, while the duty cycle can be more extended, $t_{\rm duty}\sim 10t_\phi\sim 10^{4-5}$\,yr, with $M_\bullet\sim 10^{6-8}\,M_\odot$, which circumscribes the class of Seyfert galaxies. 

Important question to address is how do thermodynamic properties of the disk affect the growth of the gravitational $m=2$ instability. Considerable effort has been invested in addressing this question. Results presented earlier in this section show a remarkable indifference of this instability to gas thermodynamics. Unless the gas is able to cool down and trigger the Jeans instability and fragmentation in the disk, the $m=2$ instability will develop first. However, development of turbulence, and possibly supersonic turbulence \citep{choi13}, in addition to the weakening of local instability due to disk thickening (Eq.\,4), removes this obstacle and allows the global instability to develop first \citep[e.g.,][]{begelman09}. Most importantly, unlike the Jeans instability which can be stabilized by long cooling times \citep[e.g.,][]{cossins09}, the global instability is indifferent to this property. 

Additional important point is that the $Q$ parameter describes the growth of axisymmetric modes, and it is frequently and unjustifiably used to analyze the non-axisymmetric modes in general, and bar-like modes in particular. Already in the linear regime it is clear that gaseous (and stellar) disk is much more unstable to non-axisymmetric modes for treatable case of tightly wound perturbations \citep[e.g.,][]{hohl71,roberts79}. Apparently the confusion comes from the dispersion relation, which, for tightly wound perturbations, becomes independent of the Fourier mode $m$. 

Three main processes can potentially lead to the gas accumulation in the galaxy centers, namely, the cosmological influx along the filaments and diffuse accretion onto the galactic disk, gravitational torques associated with mergers and close encounters, and the redistribution of gas within the galactic disk by single and nested bars. The bars themselves can be triggered by spontaneous instability discussed above or by interactions with galaxies and DM substructure \citep[e.g.,][]{noguchi88,heller07,romano08,shlosman13,bi22}. 

 \section{Discussion}
\label{sec:discuss}

Careful analysis of stellar and gas contributions in the central parsec of AGN confirms that the central SMBH dominates the region, and that megamasers in accretion disks of AGN with $M_\bullet\sim {\rm few}\times 10^6 - {\rm few}\times 10^7\,M_\odot$ participate in the Keplerian rotation. This excludes the possibility that accretion disks are globally self-gravitating. Locally self-gravitating disks, potentially Jeans unstable, do not alter the rotation curve. Of course, a larger reservoir of gas can exist and has been observed well outside the sphere of influence of the central SMBHs, on scales of 10--100\,pc and beyond. They exist either in a relaxed rotational state, e.g., NGC\,4261 showing about $10^9\,M_\odot$ circumnuclear disk \citep{sawada22}, in compact nuclear resonance rings \citep{comeron10}, or in a turbulent irregular non-equilibrium state, e.g., in NGC\,6240 during an advanced merger \citep{engel10} --- all these provide a potential feed for sub-pc accretion disks. 

Because, in principle, no limits exist on the masses of accretion disks, why do massive self-gravitating accretion disks have not been observed in AGN so far? In this work, we propose an answer to this question --- such disks are unstable to breaking their symmetry, and form gaseous bars that collapse on the center in a few rotation times, which, for $M_\bullet\sim 10^{6-8}\,M_\odot$, can be estimated conservatively as $t_\phi\sim 10^{3-4}$\,yr. 

Is this timescale related to the duty cycle of AGN which can be a factor of $\sim 10$ longer, i.e., $t_{\rm duty}\sim 10t_\phi$? Our estimate of $t_{\rm duty}$ is short compared to the SMBH characteristic growth time of $\sim 10^{7-9}$\,yr, where the lower limit comes from the detected SMBHs at $z\sim 7-10$, while the upper limit corresponds to the secular growth of the SMBHs. Observational estimate of the AGN duty cycle based on the time lag between the appearance of the central X-ray source and the photoionization time of the parent galaxy results in $t_{\rm duty}\sim 10^5$\,yr \citep{schawinski15}. 

Such disks can form multiple times, but their characteristic formation time depends on the mechanism which is responsible for the inflow, $\sim 10^{7-8}$\,yr if supplied by the processes inside $\sim 1$\,kpc, $10^{8-9}$\,yr if fed from galactic or circumgalactic scales.  Accepting the shortest rebuilding timescale of $\sim 10^7$\,yrs, the fraction of $t_\phi/10^7$\,yr is $\sim 10^{-3}$. It means that the probability of detection of globally self-gravitating disks in megamasering AGN is very small, much smaller than the current sample of these objects. Hence, their detection should focus rather on possible consequences of damping gas onto the SMBH, which can trigger either a starburst or strong wind, or both.  If a non-Keplerian disk fragments and forms stars before the collapse, the rotation curve would still remain non-Keplerian, just some of the mass would be contributed by the newborn stars. However, what we observe is $M_{\rm disk}(<r) << M_\bullet$, which means that the mass has been moved (inwards or outwards). 

If the globally self-gravitating disks collapse on the center, the gas can either join the SMBH or be (partly) expelled during nuclear starbursts. Rapid collapse will be characterized by a high accretion rate when measured close to the SMBH --- therefore, one can expect these AGN to shine close to the Eddington limit in the latter phase of the process. Under these conditions a strong outflow can develop and the AGN will appear as highly obscured objects. If dust survives the collapse, winds driven by radiation pressure on dust particles which provide an efficient coupling with radiation will be triggered. Such dusty outflows have been observed recently from obscured quasars, but lower luminosity AGN may host such outflows as well, and this can be related to transition from type\,2 to type\,1 AGN. Furthermore, one can speculate that globally self-gravitating accretion disks can form around non-masering AGN and will be subject to gravitational collapse as well. 

Both magnetic and gravitational torques are very efficient in extracting the angular momentum from the gas. Because they provide a nonlocal coupling in the accretion flow, their action can be formally described with an effective $\alpha_{\rm ss}$-viscosity, i.e., $\alpha_{\rm ss} > 1$. 

Hyper-magnetized accretion disks have been recently invoked to provide the dominant viscosity, requiring much lower surface density than self-gravitating disks \citep{hopkins26}. But this comparison was made with {\it locally} self-gravitating disks --- fragmentation in these disks can be avoided if the cooling time is longer than the orbital time. Globally self-gravitating disks can circumvent the local instabilities, and be subject to a short lifespan, during which they are very efficient in damping their mass on the central SMBH --- an alternative bypassed in \citet{hopkins26}.

While magnetic fields are probably important in AGN, direct observational evidence for this is scarce. The $B$-fields are usually related to AGN jets: to synchrotron linear/circular polarization and Event Horizon Telescope (EHT) polarimetric imaging \citep{park22}, and to transverse Faraday rotation gradients \citep{gabuzda18}. These fields relate also to recently discovered magnetic reconnection event in AGN corona using XRISM \citep{vardi26}.   

While we don't have observational evidence for globally self-gravitating disks in AGN at present, no theoretical arguments exist against them either. Moreover, a number of arguments support them, namely,

\begin{itemize}

\item Keplerian accretion disk masses in AGN can be estimated from their geometrical thickness using $M_{\rm disk}/M_\bullet\sim h/r$, which leads to a narrow range of $M_{\rm disk}\sim (0.01-0.1) M_\bullet$ observed in masering disks. There is no explanation to why the observed masering disk masses are confined to such a narrow mass range. Furthermore, it is the upper limit of $M_{\rm disk}/M_\bullet\sim 0.1$ observed in masering disks which is puzzling. A simple explanation is that more massive disks do form but are dynamically unstable and short-lived.

\item While hyper-strong magnetic fields can indeed solve the issue of fragmentation in accretion disks, gravitational torques can solve this issue as well. 

\item The SMBH seeds apparently do not form from stellar mass black hole remnants. Currently, the most plausible path to form these seeds, especially at high-$z$, is direct collapse within DM halos. Under these conditions, the massive seeds of $\sim 10^{5-6}\,M_\odot$ can form either via core collapse in supermassive stars \cite[e.g.,][]{begelman10} or via evolution of a self-gravitating disk \citep{begelman09} --- in both cases formation of a self-gravitating disk is unavoidable. 
 
\end{itemize}

Although it is usually assumed that masering disks are geometrically-thin, the possibility that the maser sources can lie in a thin surface layer above a cold massive disk has been explored \citep{modjaz05} --- an option discussed also by \citet{herrnstein05} and rejected. In this case, the mass of the underlying disk must lie in a narrow mass range, namely, being massive, on one hand, but not modifying the rotation curve, on the other hand. We discard this possibility as being too artificial. This conclusion does not change if the vertical support in the disk, such as due to turbulence and/or magnetic fields, is considered \citep[e.g.,][]{wallin98,begelman09,hopkins26}. 

Extrapolating the accretion rate in the masering region of NGC\,4258 to smaller radii, where advection-dominated accretion flow (ADAF) has been considered, brings up a discrepancy factor of $\sim 250$ in the value of $\alpha$ \citep[e.g.,][]{lasota96}. This can point to accretion rate being sharply different in various regions of the flow, which is entirely possible, if the accretion starts to rebuild from larger radii after the avalanche.

An interesting option which can potentially modify the onset of the bar instability and the catastrophic inflow is the centrifugally-driven MHD wind of clouds in the broad emission-line region (BLR) \citep{emmering92}, which has been used to discuss the cloud dynamics providing the essence of the observed NIR tori \citep{elitzur06}. Such a wind was proposed to replace the continuous masering disk \citep{kartje99}. 

The MHD wind is efficient in extracting the angular momentum from the disk, but is not required to be associated with an intensive mass outflow. So, in principle, one does not anticipate that this wind can reduce substantially the underlying disk mass. Rather, it is possible that it can supply an effective viscosity in the disk, driving a more efficient inflow.

In summary, the absence of globally self-gravitating accretion disks within the sphere of influence of central SMBHs can be explained by their short lifetime, being subjected to bar instability, and leaving behind remnant disks with $M_{\rm disk} << M_\bullet$. The end product of this sub-parsec instability and associated processes can have a profound effect on fueling activity in the galactic centers, and provide some insight into growth of the SMBHs.     

\begin{acknowledgments}
IS is grateful to Mitch Begelman for numerous discussions on the subject of self-gravitating disks, and acknowledges the usage  of Expanse CPU at San Diego Supercomputer Center (SDSC) through allocation PHY260073 from the ACCESS program, which is supported by NSF grants 2138259, 2138286, 2138307, 2137603, and 2138296 \citep{boerner23}, and the usage of the University of Kentucky Morgan Computing Cluster. 
\end{acknowledgments}

\bibliography{paper}{}
\bibliographystyle{aasjournal}

\end{document}